\documentclass[conference]{IEEEtran}
\IEEEoverridecommandlockouts

\usepackage{booktabs}
\usepackage{xcolor,colortbl}
\usepackage{color}
\usepackage{multirow}
\usepackage[T1]{fontenc} 
\definecolor{lightgray}{rgb}{0.83, 0.83, 0.83}
\usepackage{subfigure}
\usepackage[utf8]{inputenc}
\usepackage{graphicx, wrapfig,  setspace, booktabs, xcolor} 
\usepackage{graphicx}
\usepackage{amsmath}
\usepackage[version=4]{mhchem}
\usepackage{siunitx}
\usepackage{longtable}
\usepackage{tabularx}
\newcommand{\linebreakand}{%
  \end{@IEEEauthorhalign}
  \hfill\mbox{}\par
  \mbox{}\hfill\begin{@IEEEauthorhalign}
}

\usepackage{listings}

\usepackage{xcolor}

\definecolor{codegreen}{rgb}{0,0.6,0}
\definecolor{codegray}{rgb}{0.5,0.5,0.5}
\definecolor{codepurple}{rgb}{0.58,0,0.82}
\definecolor{backcolour}{rgb}{0.95,0.95,0.92}

\lstdefinestyle{mystyle}{
  backgroundcolor=\color{backcolour}, commentstyle=\color{codegreen},
  keywordstyle=\color{magenta},
  numberstyle=\tiny\color{codegray},
  stringstyle=\color{codepurple},
  basicstyle=\ttfamily\footnotesize,
  breakatwhitespace=false,         
  breaklines=true,                 
  captionpos=b,                    
  keepspaces=true,                 
  numbers=left,                    
  numbersep=5pt,                  
  showspaces=false,                
  showstringspaces=false,
  showtabs=false,                  
  tabsize=2
}

\usepackage{cite}
\usepackage{hyperref}
\usepackage{amsmath,amssymb,amsfonts}
\usepackage{algorithmic}
\usepackage{graphicx}
\usepackage{textcomp}
\usepackage{xcolor}
\def\BibTeX{{\rm B\kern-.05em{\sc i\kern-.025em b}\kern-.08em
    T\kern-.1667em\lower.7ex\hbox{E}\kern-.125emX}}
\begin{document}

\title{Efficient Training of Transfer Mapping in Physics-Infused Machine Learning Models of UAV Acoustic Field
}

\author{\IEEEauthorblockN{Rayhaan Iqbal\IEEEauthorrefmark{1}, 
Amir Behjat\IEEEauthorrefmark{2}, 
Revant Adlakha\IEEEauthorrefmark{3}, 
Jesse Callanan\IEEEauthorrefmark{4}, 
Mostafa Nouh\IEEEauthorrefmark{5}, 
Souma Chowdhury\IEEEauthorrefmark{6}}
\IEEEauthorblockA{\textit{Department of Mechanical \& Aerospace Engineering} \\
\textit{University at Buffalo}\\
Buffalo, NY, 14260\\
Email: \IEEEauthorrefmark{1}rayhaani@buffalo.edu,
\IEEEauthorrefmark{2}amirbehj@buffalo.edu,
\IEEEauthorrefmark{3}revantad@buffalo.edu,
\IEEEauthorrefmark{4}jessecal@buffalo.edu,
\IEEEauthorrefmark{5}mnouh@buffalo.edu,\\
\IEEEauthorrefmark{6}soumacho@buffalo.edu}}

\maketitle
\begin{abstract}
Physics-Infused Machine Learning (PIML) architectures aim at integrating machine learning with computationally-efficient, low-fidelity (partial) physics models, leading to improved generalizability, extrapolability, and robustness to noise, compared to pure data-driven approximation models. End-uses of PIML include, but are not limited to, model-based optimization and model-predictive control. Recently a new PIML architecture was reported by the same authors, known as Opportunistic Physics-mining Transfer Mapping Architecture or OPTMA, which transfers the original inputs into latent features using a transfer neural network; the partial physics model then uses the latent features to generate the final output that is as close as possible to the high-fidelity output. While gradient-free solvers and back-propagation with supervised learning (where optimum labels are pre-generated) have been used to train OPTMA, that approach remains computationally inefficient, and challenging to generalize across different problems or exploit state-of-the-art machine learning architectures. This paper aims to alleviate these issues by infusing the partial physics model inside the neural network, as described via tensors in the popular machine learning framework, PyTorch. Such a description also naturally allows auto-differentiation of the partial physics model, thereby enabling the use of efficient back-propagation methods to train the transfer network. The benefits of the upgraded OPTMA architecture with Automatic Differentiation (OPTMA-Net) is demonstrated by applying it to the problem of modeling the sound pressure field created by a hovering unmanned aerial vehicle (UAV). Ground truth data for this problem has been obtained from an indoor UAV noise measurement setup. Here, the partial physics model is based on the interference of acoustic pressure waves generated by an arbitrary number of acoustic monopole sources. Case studies show that OPTMA-Net provides generalization performance close to, and extrapolation performance that is 4 times better than, those given by a pure data-driven model.
\end{abstract}


\section{Introduction}
\subsection{Physics-Infused Hybrid Modeling}

Data driven models are observed to be utilized frequently for speculating the behavior of complex systems in various domains such as biological \cite{shukla2018breast}, mechanical \cite{bataineh2016neural,garimella2017neural}, robotics \cite{pfeiffer2017perception,shao2019unigrasp,rajeswaran2017learning}, and energy forecasting~\cite{ghassemi2017optimal} systems. In certain cases, wherein a “full physics” (high fidelity) model is unavailable or is computationally prohibitive to evaluate, a “partial physics” (lower fidelity) model can be used for prediction. Competitive prediction \cite{an2015practical,artun2017characterizing} is usually observed with the use of Pure data-driven models. However, they underperform at extrapolating and generalizing \cite{haley1992extrapolation,neal2012bayesian}, especially when trained with small or sparse datasets \cite{solomatine2009data}. This can be attributed to a lack of adherence to even basic physics laws; they exhibit challenging-to-interpret black-box behavior. They also exhibit sensitivity to noisy data \cite{karimi2020deep}.

With a recent review of such methods reported by Rai and Sahu \cite{hybrid-review-rai}, hybrid machine learning or hybrid surrogate modeling architectures that combine computationally efficient partial physics models with purely data-driven models in some way have been proposed as one of the solutions to the above-stated issues \cite{atamuradov2017prognostics,mehmani2015adaptive,nascimento2019fleet}.
Although there are multiple architectures available for hybrid machine learning models, Javed\cite{javed2014robust} classified them into two major sub-architectures -- the serial \cite{narendra1990identification,young2017physically,nourani2009combined,singh2019pi} and parallel sub-architectures \cite{javed2014robust,cheng2009fusion,karpatne2017physics}. The serial sub-architecture either sets the data-driven model in a series with the partial physics model or utilizes the data-driven model to tune the partial physics model parameters. 
This paper focuses on hybrid architectures that employ artificial neural networks (ANN) as their machine learning (ML) component. Our primary contribution lies in developing a new hybrid architecture that aims to: i) better utilize the partial physics model's computational efficiency and potential input-space connections with the full physics model, ii) mitigate the black-box nature of the overall predictions made by the hybrid model and iii) introduce automatic differentiation for back-propagation while training the machine learning model for better computational performance. Overall, partial physics (despite being computationally frugal) is under-exploited during the training process in most existing hybrid modeling paradigms. It is often only evaluated w.r.t. the inputs in the sparse full physics sample data. To address these issues, a novel hybrid architecture was introduced by Behjat et al. \cite{behjat2019learning} which uses partial physics as an additional node in the model. This model was called "Opportunistic Physics-mining Transfer Mapping Architecture" or OPTMA, and it was used in a dynamics prediction problem, for UAV flight prediction under gust \cite{behjat2020physics}; and in a CFD simulation modeling problem, to explore bio-inspired surface riblets for drag Reduction \cite{ghassemi2020physics}. While OPTMA outperformed data-driven and classical hybrid architectures, it was challenging to train it in a gradient-based way without generating the labels. For the dynamics prediction problem, a PSO \cite{chowdhury2013mixed} based method was used to train OPTMA, and for the airfoil design problem, supervised training was used. Both of these methods have certain limitations. This paper proposes a new architecture (OPTMA-Net) which infuses the partial physics model inside the neural network structure, with automatic differentiation used to train the model. This would be advantageous both computationally and in terms of prediction accuracy of the model, while having straightforward loss functions like mean square error. Below, we provide a summary of the transfer concept underlying the OPTMA architecture, followed by a review of automatic differentiation (AD) implementations in machine learning.

\subsection{Transfer Mapping Concept}

The OPTMA architecture can be used to model the behavior of a dynamic system in the form of a regression problem:
$\texttt{final\_state} = f(\texttt{init\_state,control\_input,elapsed\_time})$. 
Primary applications include serving as the plant model in control problems or the state transition model in reinforcement learning problems \cite{kiumarsi2017optimal,raissi2019physics}.
In this approach, a transfer mapping model is trained, to output transfer features (i.e., $f_{\texttt{OPTMA}}: X\rightarrow \chi_T$), which are fed to the partial physics model (i.e., $f_{\texttt{PP}}(\chi_T)$), based on the original input.
A new loss function is designed to account for the error between the high-fidelity, full physics output or the ground truth ($F_{\texttt{M}}$) and the predictions of the partial physics model operating on the transferred features ($F_{\texttt{PP}}(\chi_T)$). This error, 
$E = \left|F_{\texttt{PP}}(\chi_T) - F_{\texttt{M}}(X)\right|$), is typically squared and aggregated over our sparse training data set, i.e., $\forall\left(X,F_{\texttt{M}}(X)\right)\in D_{tr}$). 

\subsection{Automatic Differentiation in Training Machine Learning Models}

Gradient-based training of artificial neural networks involves back-propagating the gradients of its parameters (weights and biases) over the loss function defined. These are needed to optimize the parameters of the network using an optimization algorithm. A few common optimization algorithms used in machine learning are Gradient Descent, Stochastic Gradient Descent (SGD), and Adaptive Moment Estimation (ADAM). The most commonly faced challenge in gradient-based training is the differentiation over the network, especially when the loss function is complex or when large networks are used. Numerical Differentiation and Symbolic differentiation are two traditional methods that are used for evaluating gradients. Numerical differentiation is an easy method to implement. However, it is accompanied by rounding or truncation errors which hamper its performance. This leads to sub-optimal results. The central difference can be cited as one of the methods of numerical differentiation. Symbolic differentiation is better in the aspect of rounding error. However, the length of the representation of the resulting derivative expressions increases rapidly with the number $n$ of independent variables \cite{bischof2000computing}. Automatic differentiation is seen as the modern solution to computing gradients efficiently. Reference \cite{baydin2018automatic} gives a detailed comparison between these three differentiation techniques. 

Automatic Differentiation or Algorithmic Differentiation (AD) is one of the most fundamental and efficient tools for back-propagation in today's famous machine learning libraries like PyTorch and TensorFlow. Automatic differentiation is a set of techniques used to numerically evaluate the derivative of a function implemented within a computer program \cite{yuan2020application}. Automatic differentiation has two modes; Forward-mode AD and Reverse-mode AD. Forward-mode AD basically starts its evaluations (chain rule) at the function's inputs whose gradient is to be evaluated. Reverse-mode AD begins evaluations at the output of the function whose gradient is to be evaluated. Reference \cite{baydin2018automatic} elaborates on the two modes of AD and also the various available implementations of AD. This paper uses PyTorch's Automatic Differentiation while training the ANN-based transfer model in OPTMA-Net. The codes for implementation of OPTMA-Net for the case study are provided in a public repository\footnote{\url{https://github.com/adamslab-ub/OPTMA-Net.git}}. 

The remainder of this paper is structured as follows: the next section describes the architecture of OPTMA and its training process. Section \ref{sec:NumProb} shows the performance of OPTMA-Net on numerical toy problems. Section \ref{sec:Case Study} presents the case study taken for this paper and the description of its partial physics model. Sections \ref{sec:testing} and \ref{sec:resultConc} highlights the testing, results and conclusion of this paper. The problem statement is the determination of the acoustic field generated by a hovering unmanned aerial vehicle.



\section{OPTMA-Net}
\subsection{Previous Framework}
\begin{figure*}[h]
\centering
\includegraphics[keepaspectratio=true,scale=0.45]{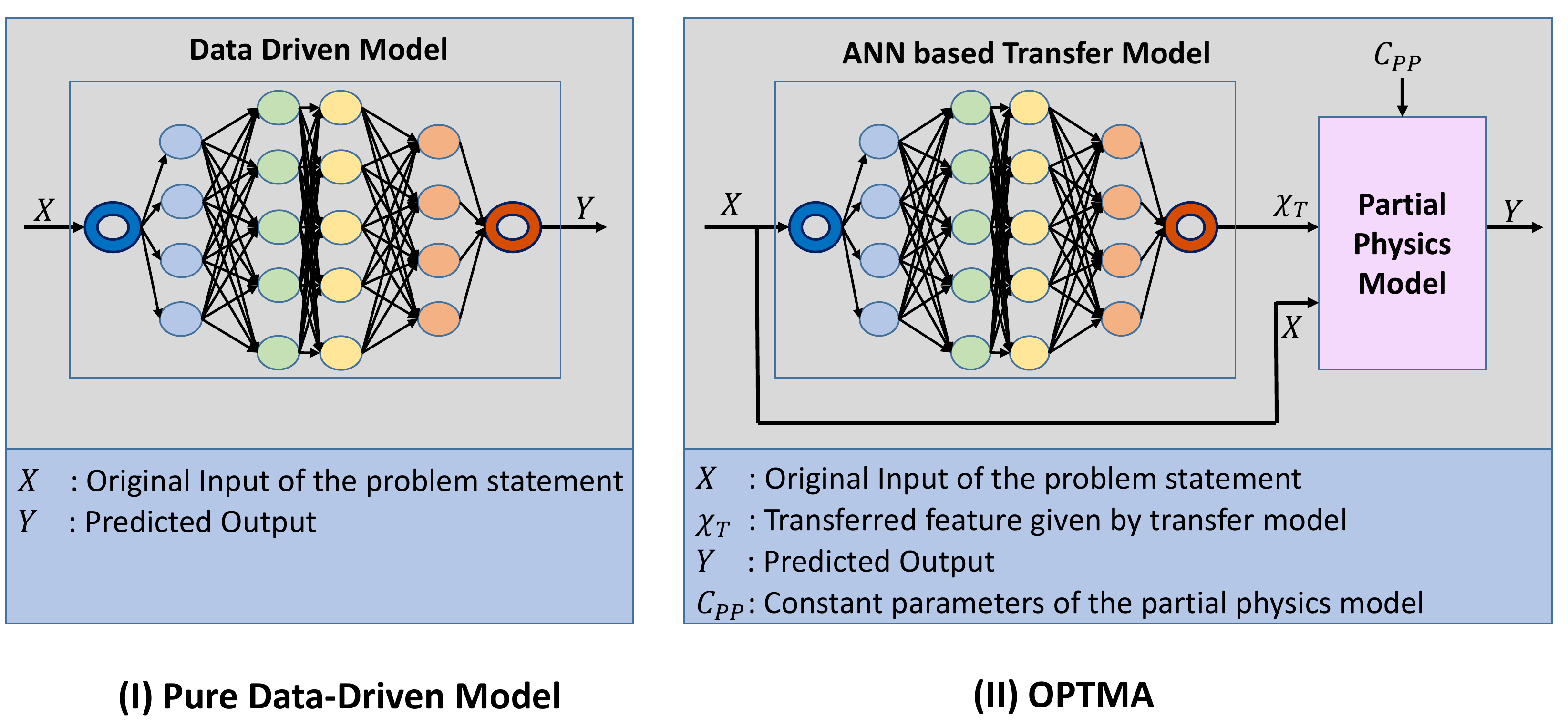}
\caption{Model Architecture: (I) Pure Data Driven Model; (II) Previous OPTMA Model.}
\label{fig:Architecture}
\end{figure*}

Figure \ref{fig:Architecture} (I) shows a standard artificial neural network that is useful for providing inexpensive and tractable approximations of the actual system response in engineering analysis and design activities. However, due to the black box nature of such models, they fail to adhere to the basic physics laws of the actual system and hence perform poorly while generalizing and extrapolating. Figure \ref{fig:Architecture} (II) shows the architecture of the    OPTMA model, which utilizes the partial physics model to ensure adherence to the physics laws of the system. It takes the input of the problem statement in the transfer model. It gives as output the optimum transferred parameters needed by the computationally inexpensive partial physics model. This results in a physics-aware surrogate model with better generalizing and extrapolating capabilities. However, training such networks can be challenging. This is because the loss in the model would have to be back-propagated over the partial physics model as well as the ANN based Transfer Model. Previously, Particle Swarm Optimization has been used to train OPTMA \cite{behjat2020physics}. This is however, computationally very expensive and would not be viable for complex physics-based models. To solve this problem of back-propagation, OPTMA-Net is proposed in this paper, which utilizes PyTorch's AD for back-propagation.

\subsection{OPTMA-Net Framework}
\begin{figure*}[h]
\centering
\includegraphics[keepaspectratio=true,scale=0.45]{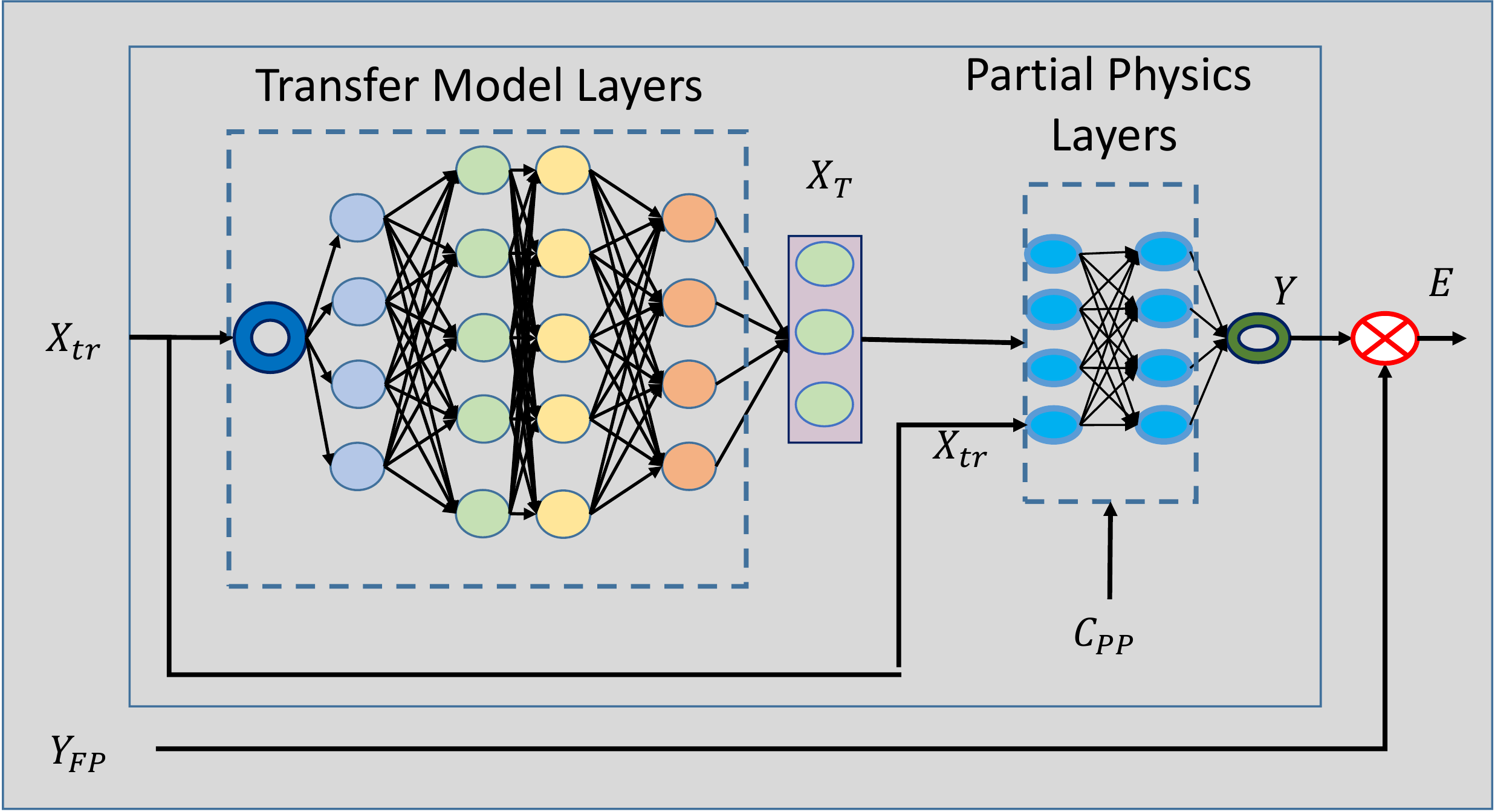}
\caption{Model Architecture: OPTMA-Net}
\label{fig:OPTMA-Net}
\end{figure*}

In this paper, we propose a model which utilizes a partial physics model in a neural network architecture, along with an efficient training procedure. To accomplish this, the partial physics is included inside the ANN based Transfer Model itself as shown in \ref{fig:OPTMA-Net}. It can be thought of as the last layer or node of the Transfer Model. These are shown as Partial Physics Layers in \ref{fig:OPTMA-Net}, and represented the exact partial physics function, written in terms of PyTorch compatible tensors. Utmost care must be taken to ensure that these layers are compatible with PyTorch's back-propagation. This enables using simple loss functions like Mean Square Error (MSE) and seamless training through AD. This makes the model's training computationally fast and gives a unified model without passing the output through the partial physics model.

\subsection{PyTorch Modelling}
\subsubsection{ Automatic Differentiation}
Since OPTMA-Net is a PyTorch based model, it utilizes automatic differentiation (AD) for training the model. We briefly describe AD as used in back-propagation. A detailed description of the working of automatic differentiation can be found in \cite{su1997automatic, griewank2008evaluating, griewank2014automatic}. AD relies on the ability to decompose a program into a series of elementary operations (primitives) for which the derivatives are known and to which the chain rule can be applied \cite{van2018automatic}. Currently, there are two famous implementation techniques in AD: \textit{Operator Overloading} (OO) and \textit{Source Code Transformation} (SCT). To better understand the complete working behind computing gradients, the readers can refer to \cite{bischof2000computing}. While both implementations have their advantages and disadvantages, OO tends to be easier to implement. PyTorch's automatic differentiation has the OO implementation \cite{paszke2017automatic}. The loss function is back-propagated over the Partial Physics Layers as well as the Transfer Model automatically.

\subsubsection{Partial Physics layer}
For PyTorch's AD to be compatible with the OPTMA-Net model, all layers and mathematical functions must be defined based on tensors \cite{ketkar2017introduction}. To write the custom layers constituting the partial physics model, we define the function evaluations in terms of torch tensors. Thus we can write any compatible set of mathematical operations in terms of tensors to represent a partial physics model as an incorporated layer of OPTMA-Net. The implementation for the partial physics acoustic model, as used in our case study, is shown in Section \ref{ssec:partialphysicspytorch}.

\subsubsection{Extract intermediate layer values}

Many partial physics models have input features (the transferred feature) as not just constants for the model but might have physical relevance with respect to the problem statement. For this reason, it might be needed to analyze the transferred inputs generated by a trained OPTMA-Net model. 

For this reason, we define the method used to retrieve $\chi_T$, produced by the transfer model section of OPTMA-Net. Essentially, this is the output of the final layer used in the transfer model part of OPTMA-Net. This is retrieved by placing a hook on the final layer of the transfer model (use PyTorch's\\ \emph{\texttt{register\_forward\_hook}} on the mentioned final layer). This would give us the activation of any desired layer, where we register a hook. In this way, we can retrieve intermediate values of OPTMA-Net, which would have physical relevance to the problem statement at hand.

\subsection{Physics Layer PyTorch complexity}

The obvious question that may arise while selecting partial physics models to use for OPTMA-Net is the level of complexity of the partial physics model to be used. Obviously, if the partial physics model is itself computationally expensive, relative to a full physics model of the problem, training OPTMA-Net would be much slower. Here lies the trade-off between implementing a more accurate model in OPTMA-Net for higher accuracy vs. the added computational expense due to the increased complexity of the partial physics model. Now, suppose we wish to increase the complexity of the partial physics model (choose a more accurate implementation for the partial physics). In that case, we can picture this as a change in the structure of the partial physics layers of OPTMA-Net. For every added node, the computation time increases by the number of training samples times the number of training epochs times the number of added nodes. This is the trade-off between computation time and partial physics model complexity to be considered while setting up OPTMA-Net.


\section{Numerical Problems}\label{sec:NumProb}

The numerical problem on which OPTMA-Net is tested is the same as used by Behjat et al. \cite{behjat2020physics}, the Gramacy and Lee Problem. For testing Baselines, OPTMA-Net is compared with a pure data driven model and a sequential hybrid physics-infused model.
\subsection{Gramacy and Lee Problem}\label{ch4-sec:gram}
Just as in \cite{behjat2019learning}, OPTMA-Net will be used to predict the output for two separate full physics functions. The partial physics function would be the same for both cases. Once structured and trained, OPTMA-Net is expected to transfer the input features and compute the corresponding partial physics outputs with a minimum error compared to the full physics outputs (as described in previous sections). The description of the partial physics and full physics functions is given below.\\ 
\par\emph{Partial Physics Model:}
\begin{equation}
\small
\label{eq:Gramacy-LeePP}
\begin{aligned}
F_{\texttt{PP}}(x
) = \frac{\sin(10 \pi x)}{2x} +(x-1)^4 ; \ x \in [0.5, 2.5] \\
\end{aligned}
\end{equation}

\par\emph{Full Physics Model 1 (FP1):}
\begin{equation}
\small
\label{eq:Gramacy-Lee1}
\begin{aligned}
F_{\texttt{FP1}}(x) =& F_{\texttt{PP}}(3-x)\\
F_{\texttt{FP1}}(x) =& \frac{\sin(10 \pi (3-x))}{2(3-x)} +((3-x)-1)^4 ; \ x \in [0.5, 2.5] \\
\end{aligned}
\end{equation}

\par\emph{Full Physics Model 2 (FP2):}
\begin{equation}
\small
\label{eq:Gramacy-Lee2}
\begin{aligned}
F_{\texttt{FP2}}(x) =& F_{\texttt{PP}}((0.5 + 2 \sin\left(\pi ({x-2})/{2}\right)))\\
F_{\texttt{FP2}}(x) =& \frac{\sin(10 \pi (0.5 + 2 \sin\left(\pi ({x-2})/{2}\right)))}{2 (0.5 + 2 \sin\left(\pi ({x-2})/{2}\right))}\\
&+((0.5 + 2 \sin\left(\pi ({x-2})/{2}\right))-1)^4 ; \ x \in [0.5, 2.5] \\
\end{aligned}
\end{equation}
Where $F_{\texttt{PP}}$ is the partial physics model, $F_{\texttt{FP}}$ is the full physics model 1 and $F_{\texttt{FP2}}$ is the full physics model 2.

\subsection{Baselines}\label{ch4-sec:baseline}
\subsubsection{Pure Data Driven model}
A pure data driven model is taken as one of the baselines for comparisons. It has its benefits in being easy to implement and fast to train. Since the loss function is usually straightforward (MSE, BCE, etc.), training is not difficult. However, as mentioned before, its black box like nature often causes a lack of adherence to basic physical laws of the problem statement. For this paper, an ANN based pure data driven model is written in PyTorch. To ensure a fair comparison, its network configurations are kept as similar as that used for OPTMA-Net. This means that the number of overall layers, number of nodes per layer, maximum epochs, and batch size are kept the same. Obviously, the actual layers of the model differ from those in OPTMA-Net.

\subsubsection{Sequential Hybrid Physics Infused Model}
OPTMA-Net is also compared with a sequential hybrid physics-infused model, which used the outputs of the partial physics model as additional inputs to the neural network. The details of the sequential hybrid model can be found in \cite{callanan2021large}. This highlights the effect of using the partial physics inside the structure of the ANN, used in OPTMA-Net. 

\subsection{Modelling Parameters}
\label{ch4-sec:param}

To make it a fair comparison between all models involved, we try to have similar parameters of all the compared models. Obviously, the layer description would differ in OPTMA-Net (it consists of added custom layers representing the partial physics model). When we say model parameters, we mean the number of layers used, number of hidden nodes per layer, number of input and output nodes, learning rate, etc. The detailed parameter description for all models, used to solve both numerical problem statements, i.e. FP1 (equation\ref{eq:Gramacy-Lee1}) and FP2 (equation~\ref{eq:Gramacy-Lee2}, are given in the Table~\ref{ch4-tab:param_fp}).

\begin{table*}[t]
\begin{center}
\caption{\textbf{Model Parameters to solve FP1 and FP2}}
\begin{tabular}{ c c c c c }
\toprule
Parameters & Pure DataDriven & Sequential Hybrid Model & OPTMA-Net \\
\hline 
Num Inputs & 1 & 2 & 1\\ 
Num Outputs & 1 & 1 & 1\\
Num. Layers & 3 & 3 & 3+1\\
Nodes per layer & 50 & 50 & 50\\
Learning Rate & $1\times 10^{-5}$ & $1\times 10^{-5}$ & $1\times 10^{-5}$\\
Batch size & 10 & 10 & 10\\
Dropout & 0.1 & 0.1 & 0.1\\
Testing size & 200 & 200 & 200\\
\bottomrule
\end{tabular}
\label{ch4-tab:param_fp}
\end{center}
\end{table*}

\subsection{Results}
\label{ch4-sec:results}
This paper uses the Normalized RMSE as the error metric to compare the results of Gramacy and Lee problem. This error metric can be described as follows:
\begin{equation}
       \mathrm{RMSE} =  \sqrt{ \frac{ \sum_{n=1}^{n_{S}} (Y_i^{M} - Y_i^P)^2}{n_{S}}}
       \label{eq:RMSE}
\end{equation}

Where, $n_{S}$ are the total number of samples in the testing dataset, $Y_i^{M}$ are the ground truth values of the testing dataset, $Y_i^P$ are the model predicted values. The RMSE is calculated over the normalized predictions. Tables~\ref{ch4-tab:res_fp1} and \ref{ch4-tab:res_fp2} shows the results for FP1 and FP2 respectively. The models are tested over a fixed testing dataset of size 200. The tests are done over a variable training dataset size. OPTMA-PSO shows the results as reported in \cite{behjat2020physics}.

\begin{table*}[t]
\begin{centering}
\caption{\textbf{Normalized RMSE testing dataset results for all models to solve FP1}}
\makebox[\textwidth][c]{%
\begin{tabular}{ c c c c c }
\toprule
Training Size & Pure DataDriven & Sequential Hybrid Model & OPTMA-PSO \cite{behjat2020physics} & OPTMA-Net \\
\hline 
20 & 0.243 & 0.106 & \textbf{0.039} & 0.094\\
50 & 0.111 & 0.081 & \textbf{0.019} & 0.074\\
100 & 0.073 & 0.052 & \textbf{0.028}  & 0.049\\
200 & 0.064 & 0.041 & 0.036 & \textbf{0.022}\\
\bottomrule
\end{tabular}
}
\label{ch4-tab:res_fp1}
\end{centering}
\end{table*}

\begin{table*}[t]
\begin{centering}
\caption{\textbf{Normalized RMSE testing dataset results for all models to solve FP2}}
\makebox[\textwidth][c]{%
\begin{tabular}{ c c c c c }
\toprule
Training Size & Pure DataDriven & Sequential Hybrid Model & OPTMA-PSO \cite{behjat2020physics} & OPTMA-Net \\
\hline 
20 & 0.172 & 0.211 & \textbf{0.061} & 0.072\\
50 & 0.123 & 0.068 & 0.080 & \textbf{0.055}\\
100 & 0.086 & 0.075 & 0.088 & \textbf{0.048}\\
200 & 0.079 & 0.059 & 0.083 & \textbf{0.040}\\
\bottomrule
\end{tabular}
}
\label{ch4-tab:res_fp2}
\end{centering}
\end{table*}

\section{Case Study: Acoustic field generated by a hovering UAV}\label{sec:Case Study}

\subsection{Problem Statement and Data Collection}
\begin{figure*}[h]
\centering

\includegraphics[width=0.6\textwidth]
{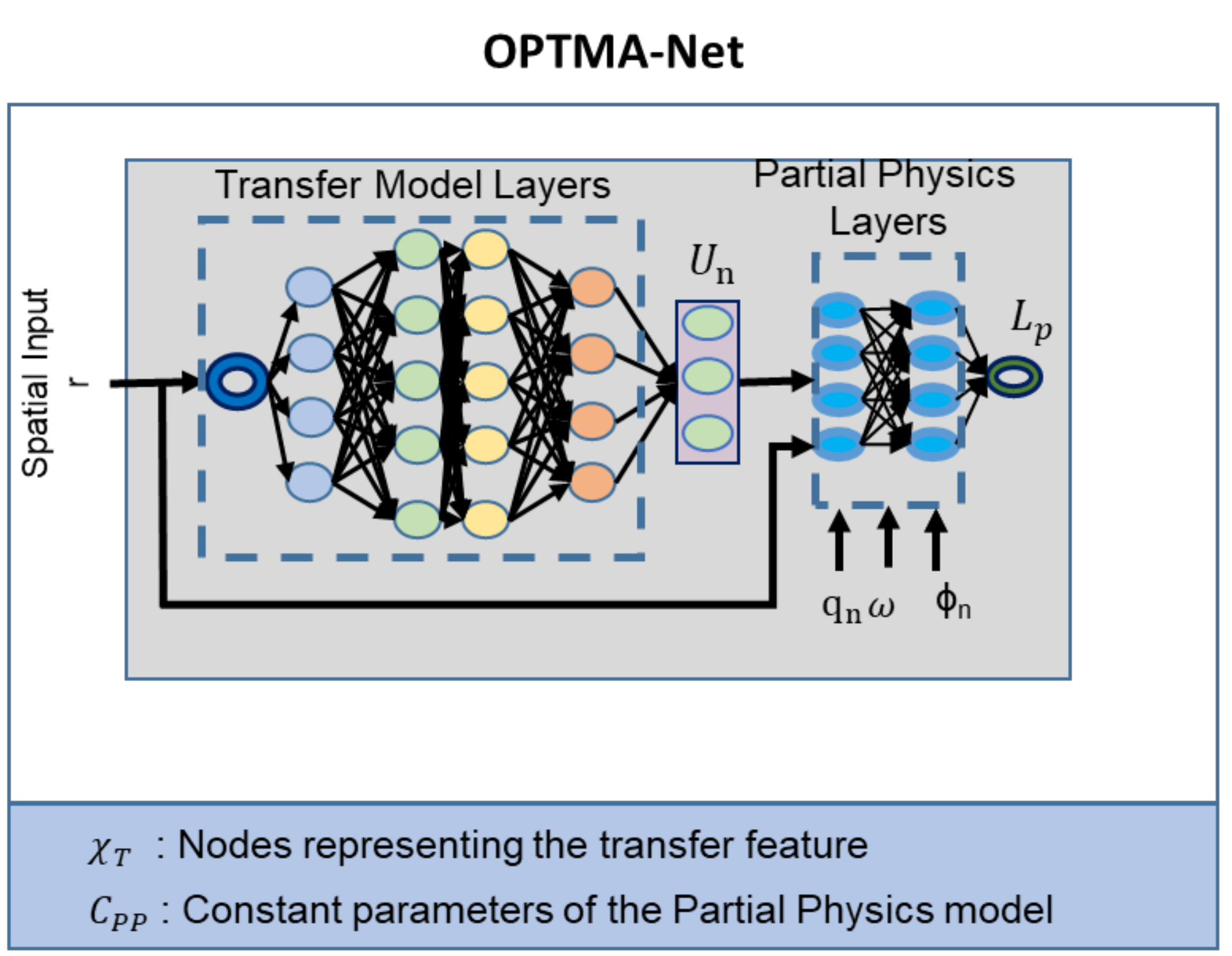}
\caption{Inputs and outputs of OPTMA-Net for transfer model and partial physics model: Acoustic field modeling }
\label{fig:inps_outps_framework}
\end{figure*}

For this case study, the motivation is to estimate the entire spatial acoustic pressure field around a hovering UAV based on a set of experimental measurements. It is of significant importance to be able to accurately estimate the noise emitted by a UAV for design, integration, and safety purposes \cite{callanan2020ergonomic}. Once an accurate model of the acoustic pressure field around the UAV is available, designers and engineers can predict the impact of the noise on nearby human agents, wildlife, and sensors or equipment. Figure \ref{fig:inps_outps_framework} illustrates the inputs and puts to transfer model and partial physics. The input to such a model is the location at which the sound pressure level is to be estimated relative to the UAV location. A previous study carried out by this group \cite{callanan2021large} performed the first large-area experimental measurements of the sound field produced by an unconstrained hovering UAV. A series of experimental measurements were carried out to measure the sound pressure level at over 1,700 individual locations around a hovering UAV in order to build a map of the acoustic field intensity as a function of space. The full results of that experiment are available \cite{callanan2021large} and some key points are highlighted here. The UAV used in that study was a DJI Phantom. A custom microphone array apparatus, called Large Aperture Scanning Microphone Array (LASMA), was constructed to measure acoustic pressure field data for a steady-state noise source over a large area. The total scan area was a rectangle of width 2.3 m and height 1.2 m. The acoustic measurements were taken with four BSWA MPA416 pre-polarized $^1/_4$" microphones (IEC 61672 Class 1, $\pm$1dB at 1 kHz, $\pm$2dB over 20 Hz-10 kHz) and a USB MC3522 DAQ. The location of the microphones and the sound source were tracked in real time with a Vicon motion capture system. The Vicon motion capture system consisted of Vicon Vantage V8 and Vero v2.2 optical capture cameras and Vicon Tracker software which is capable of tracking any object with appropriate retro-reflective markers within fractions of a millimeter. The LASMA, microphones, and UAV or speaker source were equipped with the necessary markers, and their position was tracked dynamically throughout the scans. The position data is used to map the microphone readings to the 3-dimensional space, thus constructing a full map of the sound distribution around the source. The result is a series of data points which represent samples of a scalar field distributed irregularly over a large area; each data point is the sound pressure level at some spatial location.

\subsection{Partial Physics Acoustic Model}\label{ssec:partialphysics}
A partial physics model to characterize an acoustic pressure field is presented using an arbitrary number of acoustic spherical sources in an ideal environment. These acoustic spherical sources, when tuned to optimal individual amplitude, frequency, phase, and relative position, can reconstruct the acoustic pressure field obtained from an experiment setting with high accuracy due to constructive and destructive interference. It is worth noting, that these acoustic spherical sources do not represent the location of the rotors or other noise producing sources in the UAV drone, but simply represent ideal spherical sources placed in an ideal environment to recreate the experimentally obtained pressure field. The time-independent pressure field generated by an individual acoustic spherical source is defined as
\begin{equation}
       p_n(\mathbf{r}) = \frac{{U}_{n}}{{r}_n} \cos(\omega {r}_n /c + \phi_n)
       \label{press_monopole}
\end{equation}
where $\mathbf{r}$ is the position vector of the field point relative to the origin, $\mathbf{r}_n = \mathbf{r} - \mathbf{q}_n$ with $\mathbf{q}_n$ as the position vector of the $n^{\mathrm{th}}$ acoustic source relative to the origin, and ${r}_n = |\mathbf{r}_n|$ is the Euclidean distance of the field point from the $n^{\mathrm{th}}$ acoustic source. The source has normalized amplitude $U_n$, angular frequency $\omega$, the speed of sound defined at STP as $c = 343 $ m/s, and $\phi$ as the phase angle. The subscript $(\bullet)_n$ is the shorthand notation for the $n^\mathrm{th}$ acoustic spherical source. The net acoustic pressure field can be computed by adding the pressure fields of all the acoustic sources as $P(\mathbf{r}) = \sum^{N}_{n = 1} p_n(\mathbf{r})$, where $N$ is the total number of acoustic sources. Finally, the sound pressure level (SPL) can be computed as $L_p(\mathbf{r}) = 20 \, \mathrm{log_{10}}[|P(\mathbf{r})|/P_\mathrm{ref}]$, where $P_\mathrm{ref} = 20 \ \mu$Pa and $|\bullet|$ is the absolute value.It is important to note, that we are using the absolute value of the acoustic pressure instead of the more traditionally used root mean squared (RMS) value.

\subsection{Partial Physics Parameters and Transfer Feature}\label{ssec:transferred feature}
The constant parameters of the partial physics model include $c$, $\omega$, and $\phi$. These act as $C_{PP}$ as shown in Figure \ref{fig:Architecture} (II). These values can either be selected arbitrarily or by a static optimization procedure. We can optimize these constant parameters, one by one, by fixing the other parameters to a random value. There are multiple ways of carrying out this optimization. This paper will aim to optimize the constant parameters before using them in the partial physics model.

The normalized amplitudes $U_n$ are taken as transfer features for this problem. An ideal transfer model maps the transferred feature for the partial physics model such that the partial physics model generates the actual outputs (i.e., the full physics experimental values). Equation \eqref{eq:transfer_vars} show the role of transfer model for the partial physics model.

\begin{equation}\label{eq:transfer_vars}
       U_n = \mathcal{TM} (r_n)
\end{equation}

where $\mathcal{TM}$ is the trained transfer model. The transfer model should be able to predict the optimum $U_n$ for any input location $\mathbf{r}_n$. The optimum $U_n$ along with the constant parameters are fed to the partial physics model to predict the sound pressure levels generated at the input location.

\subsection{Partial Physics PyTorch Layers}\label{ssec:partialphysicspytorch}
The partial physics layers in PyTorch must be written in terms of PyTorch tensors for it to be compatible with back-propagation utilizing AD. The network code for the partial physics acoustic model is shown in Listing \ref{code:network}. Lines 26-36 show the implementation of partial physics in PyTorch compatible terms. 

\lstset{style=mystyle}
\begin{figure*}
\begin{minipage}{2\columnwidth}
\begin{lstlisting}[language=Python, caption=Network implementation PyTorch,label={code:network}]
import torch
import config1 as c     #Import Partial Physics configuration 

class Fully_connected(torch.nn.Module):
    def __init__(self, D_in, D_out,config):
        super(Fully_connected, self).__init__()
        self.layers = torch.nn.ModuleList()
        H = config['hidden_layer_size']
        self.norm = torch.nn.BatchNorm1d(D_in)
        self.linear_in = torch.nn.Linear(D_in, H)
        self.dropoutp = config['dropout']
        for i in range(c.Num_layers):
            self.layers.append(torch.nn.Linear(H,H))
        self.drop = torch.nn.Dropout(p=self.dropoutp)
        self.linear_out = torch.nn.Linear(H, D_out)
        self.nl1 = torch.nn.ReLU()

    def forward(self, x):
        #Transfer Model Layers: Standard Layers of desired description
        out = self.linear_in(self.norm(x))
        for i in range(len(self.layers)):
            net = self.layers[i]
            out = self.nl1(self.drop(net(out)))
        out = self.linear_out(out)
        
        #Partial Physics Layers: Custom PyTorch compatible layers
        P = torch.zeros(out.shape[0],1,
                        dtype=torch.cfloat).to(c.device)    
        for n in range (0,4):   #Loop over N=4 monopoles
            r=torch.sqrt(torch.pow(x[:,0]-c.mono_loc[0,n],2)+
                         torch.pow(x[:,1]-c.mono_loc[1,n],2)+
                         torch.pow(x[:,2]-c.mono_loc[2,n],2))
            P[:,0]=P[:,0]+(out[:,n]*torch.cos
                        (c.kappa[n]*r+c.phi[0,n]))/r
        spl = (20* torch.log10(torch.abs(P)/c.P_ref))
        return spl  #Normalize before returning

def l2_loss(input, target):    
     loss = torch.nn.MSELoss()
     return loss(input,target)
 
def make_train_step(model,optimizer,scheduler=None):
    #One step in the training loop
    def train_step(x, y,test=False):
        a=model
        if not test:
            yhat = a(x)
            loss = l2_loss(yhat, y)
            optimizer.zero_grad()
            loss.backward()
            optimizer.step()
        else:
            a = model.eval()
            with torch.no_grad():
                yhat = a(x)
                loss = l2_loss(yhat, y)
            if scheduler:
                scheduler.step(loss)
        return loss.item()
    #Returns the function called during training
    return train_step
\end{lstlisting}
\end{minipage}
\end{figure*}


The configurations or constant parameters ($C_{PP}$ in Figure \ref{fig:Architecture}) can be imported from a separate file. It must be noted that these configurations must be torch tensors as well for compatibility with PyTorch's AD for back-propagation. Code Listing \ref{code:config} shows the configuration for OPTMA-Net. Lines 3-8 define the transfer model's parameters. Lines 11-21 describe the parameters for the partial physics model. These configurations are imported in our network as \textit{config1}. Obviously, the network can be called inside any standard training loop written in PyTorch.
\begin{figure*}
\begin{minipage}{2\columnwidth}
\begin{lstlisting}[language=Python, caption=Partial Physics Configuration PyTorch: \textit{config1}, label={code:config}]
import torch
#Transfer Model Parameters
device = torch.device('cuda:0')
D_in = 3; D_invin = 80; D_invout = 30; D_out = 4
lr = 1e-4; dropout = 0.1
batch_size = 25; epochs = 100
Hidden_layer_size = 50; Num_layers = 3
weight_decay = 0

#Partial Physics Parameters
mono_loc=torch.cuda.FloatTensor([[0.176,  -0.176,  -0.176,  0.176], 
                            [0.176,  0.176,  -0.176,  -0.176],
                            [0,  0,  0,  0]])

comp_1i = torch.tensor([[0.0 + 1j]]).to(device)
phi = torch.cuda.FloatTensor([[45,  45,  45,  45]])
P_ref = torch.cuda.FloatTensor([[20e-6]])
freq=torch.cuda.FloatTensor([[175,	175,	175,	175]])
pi = torch.acos(torch.zeros(1)).item() * 2
ang_freq = 2*pi*freq[0,:]
kappa = ang_freq/343
\end{lstlisting}
\end{minipage}
\end{figure*}

\section{Testing}\label{sec:testing}
This section carries out a thorough testing comparison between the Pure Data Driven model, the Sequential Hybrid Physics-Infused model, and OPTMA-Net. For this purpose, we create the training and testing datasets in 3 different ways: i) Percentage based splitting, ii) Quadrant based splitting, and iii) Radial based splitting. The Quadrant and Radial testing are done to measure the extrapolating capabilities of each model, whereas the percentage-based testing will measure the model's generalizability. 

\subsection{Percentage Testing}
\begin{figure*}
\centering
\includegraphics[width=0.99\textwidth]
{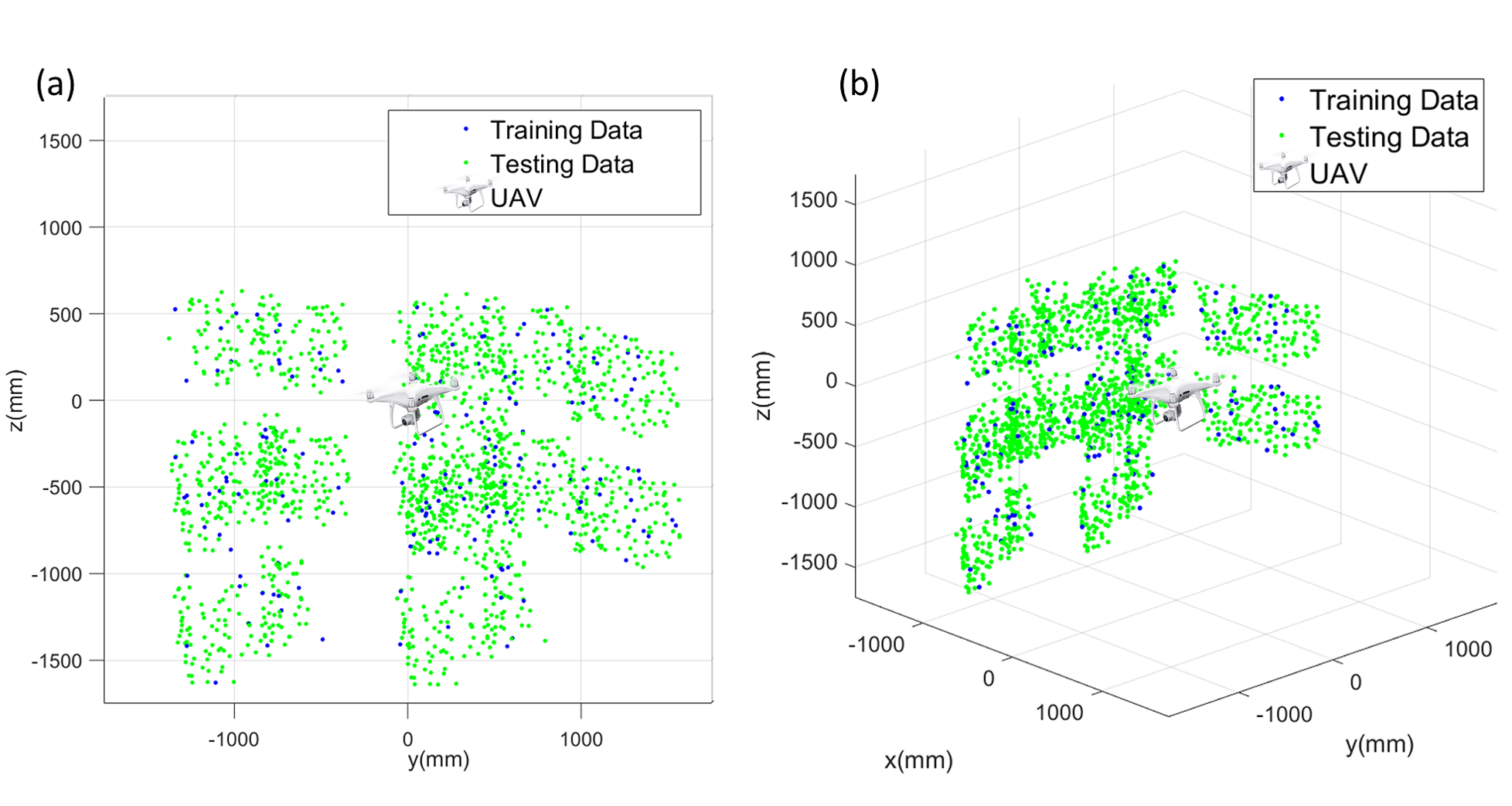}
\caption{Percentage Testing: Shows the training and testing data when using only 10\% of the entire dataset to train. (a) \textit{y-z} plane view of the training and testing data points. (b) 3D view of the training and testing data points }
\label{fig:Perc}
\end{figure*}

Here, we simply create the training dataset from different percentages of the entire dataset. The percentages vary from 10\% to 90\% of the whole dataset, i.e., the training dataset will be varied between 10\% and 90\% of the entire dataset. The same datasets are used to train all 3 models. The datasets are uniformly sampled over the spatial locations, as can be seen in Fig. ~\ref{fig:Perc}. In this figure, 10\% of the entire dataset has been used for training the models. Percentage-based testing is used to measure the generalizability of models. 

\begin{figure*}
\centering
\includegraphics[width=0.99\textwidth]
{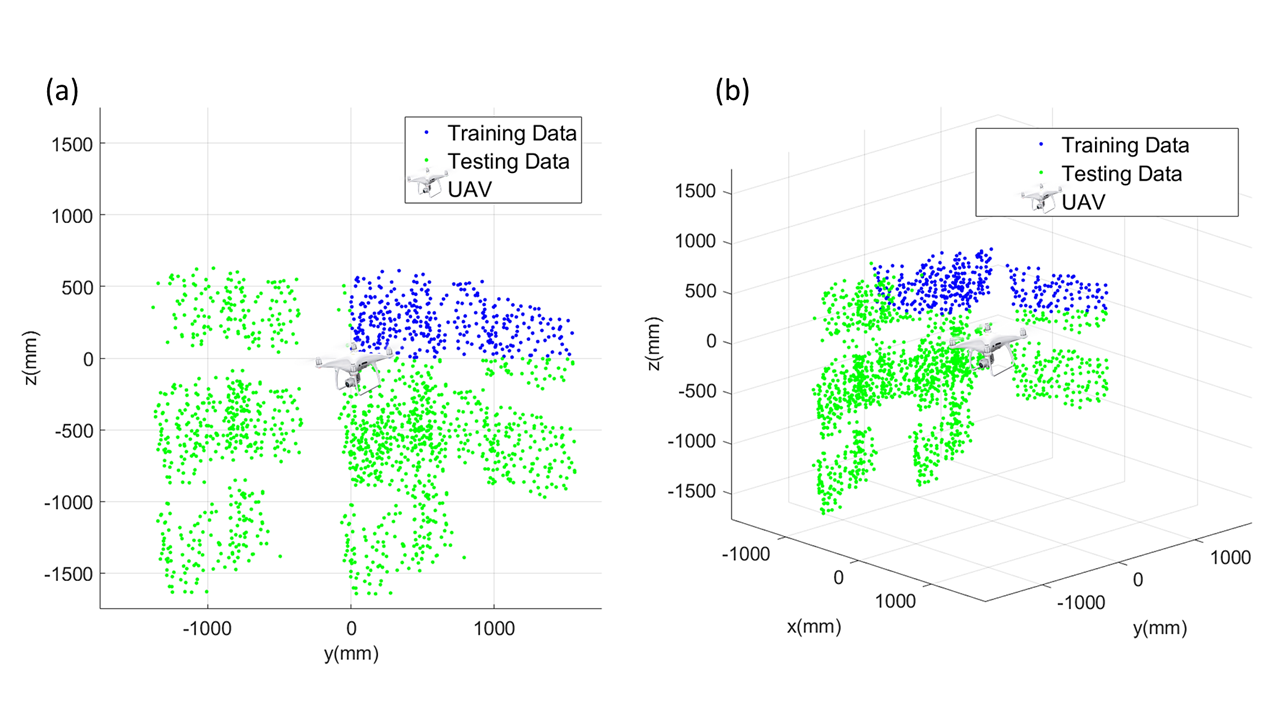}

\caption{Quadrant Testing: Shows the training and testing data used to train the models. (a) \textit{y-z} plane, showing the training and testing data points. (b) 3D view showing the training and testing points.} 
\label{fig:Quad}
\end{figure*}
%

\subsection{Quadrant Testing}
\vspace{-0.15cm}
An ideal machine learning model should be not only good at generalizing but also be capable of extrapolating. The quadrant-based testing measures the extrapolating capabilities of each model. Here, the training dataset consists of all values present in the first quadrant of the \textit{y-z} plane, and the model is tested in the remaining three quadrants of the \textit{y-z} plane. This is shown in Fig. ~\ref{fig:Quad}. 
Quadrant splitting is a formidable sampling method for acoustic problems specifically due to the potential for asymmetry in the acoustic field which is not necessarily split across quadrants. In the case of the UAV sound source specifically, it is reasonable to expect that the lower two quadrants (or four octants, and location with negative $z$-value) will exhibit higher SPL values due to the directivity of the noise produced by the spinning propellers. On the contrary, the acoustic field is symmetric in the x-y plane as the propellers are spinning with equal and constant angular velocities to hold the drone in place. In hovering mode, these propellers generate equal upward thrust to counteract the weight of the drone and as a consequence create down drag which raises the SPL values in the lower two quadrants creating an asymmetric acoustic field in x-z and y-z planes. It can also be noted, that the SPL decreases as the distance between the UAV and measurement location is increased creating a sort of radial symmetry. In other words, the SPL is inversely proportional to the distance between the the source and receiver. Therefore a model which can accurately predict the SPL value in a quadrant where no training data was present is likely accurately capturing the true physics of the problem.


\subsection{Radial Testing}
\begin{figure*}
\centering
\includegraphics[width=0.99\textwidth]
{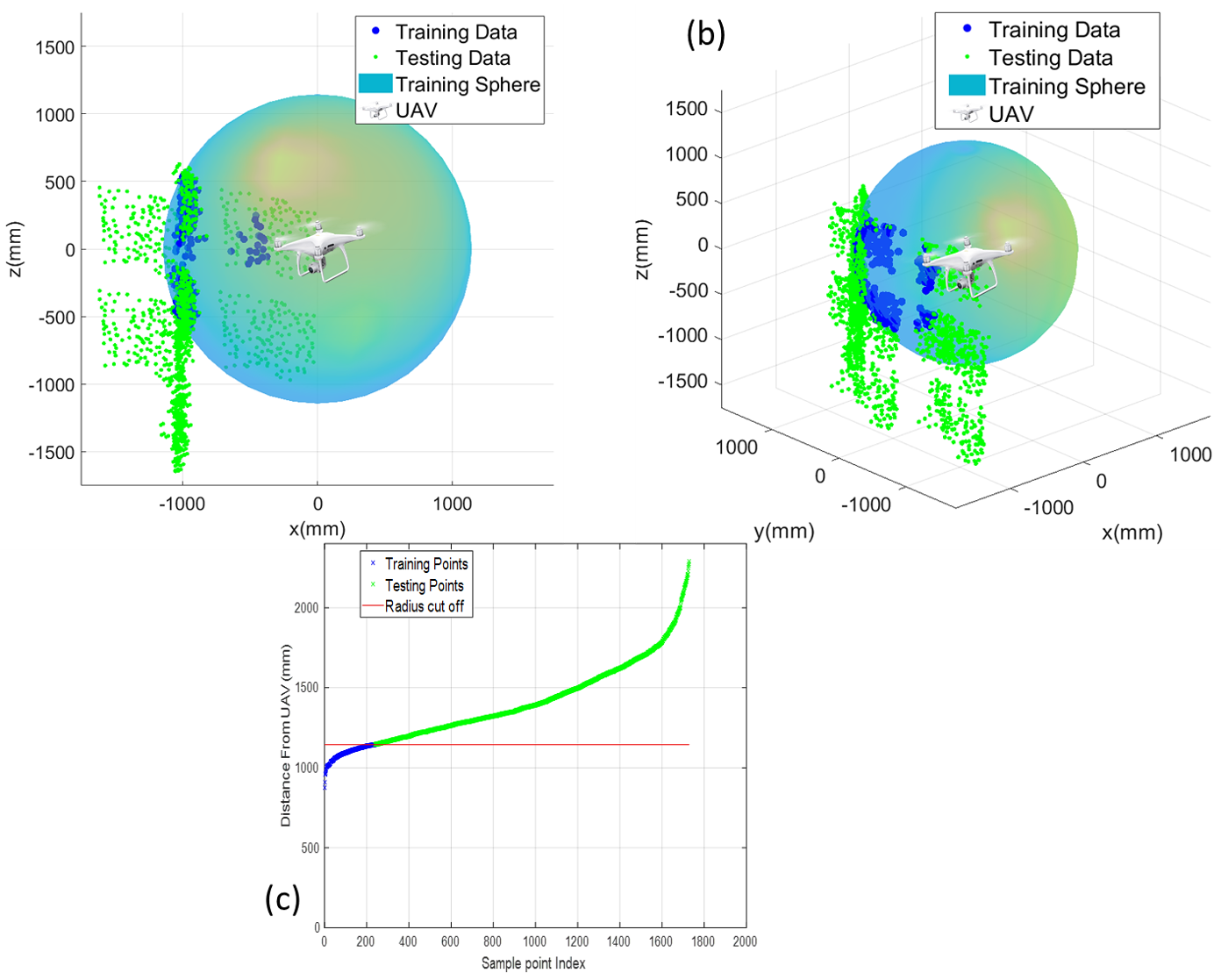}
\caption{Radial Testing: Creating training data based on radial distance of points from UAV. (a) \textit{x-z} plane view of the training and testing data along with the sphere splitting them. (b) 3D view of the training and testing data points along with the splitting sphere. (c) Plot showing the Sample Point VS Distance from the UAV for the training and testing dataset along with the radius of the splitting sphere.}
\label{fig:Radial}
\end{figure*}

In this testing, we divide the train-test dataset based on the radial distance of sample points from the UAV. The training dataset comprises all points enclosed within an imaginary sphere drawn with the UAV as its center and the radius equal to half the distance between the furthest sampled point and the UAV. This can be seen in Fig. \ref{fig:Radial}(a) and \ref{fig:Radial}(b). Radial splitting forms a dataset of only the points that are much closer in the distance to the UAV. Fig. \ref{fig:Radial}(c) shows the variation of the distance of sample points from the UAV. This test also acts as a measure of the machine learning models extrapolating capabilities.

%



\section{Results}\label{sec:resultConc}

The results presented are those carried out on the testing dataset. This paper uses the Normalized Mean Square Error (MSE) and the Relative error (RE) as the error metrics to compare the results of the acoustic field prediction problem. These error metrics can be described as follows:
\begin{equation}
       \mathrm{MSE} =  \frac{ \sum_{n=1}^{n_{S}} (Y_i^{M_N} - Y_i^{P_N})^2}{n_{S}}
       \label{eq:MSE}
\end{equation}

\begin{equation}
       \mathrm{RE_i} =  \frac{ (Y_i^{M} - Y_i^P)}{Y_i^{M}}\times100
       \label{eq:RAE}
\end{equation}
where, $n_{S}$ are the total number of samples in the testing dataset, $Y_i^{M}$ are the ground truth values of the testing dataset, $Y_i^P$ are the model predicted values. $Y_i^{M_N}$ are the ground truth values Normalized between 0-1 and $Y_i^{P_N}$ are the predictions Normalized between 0-1. The MSE is calculated over the normalized predictions. 

\begin{figure*}
\centering
\includegraphics[keepaspectratio=true,scale=0.79]{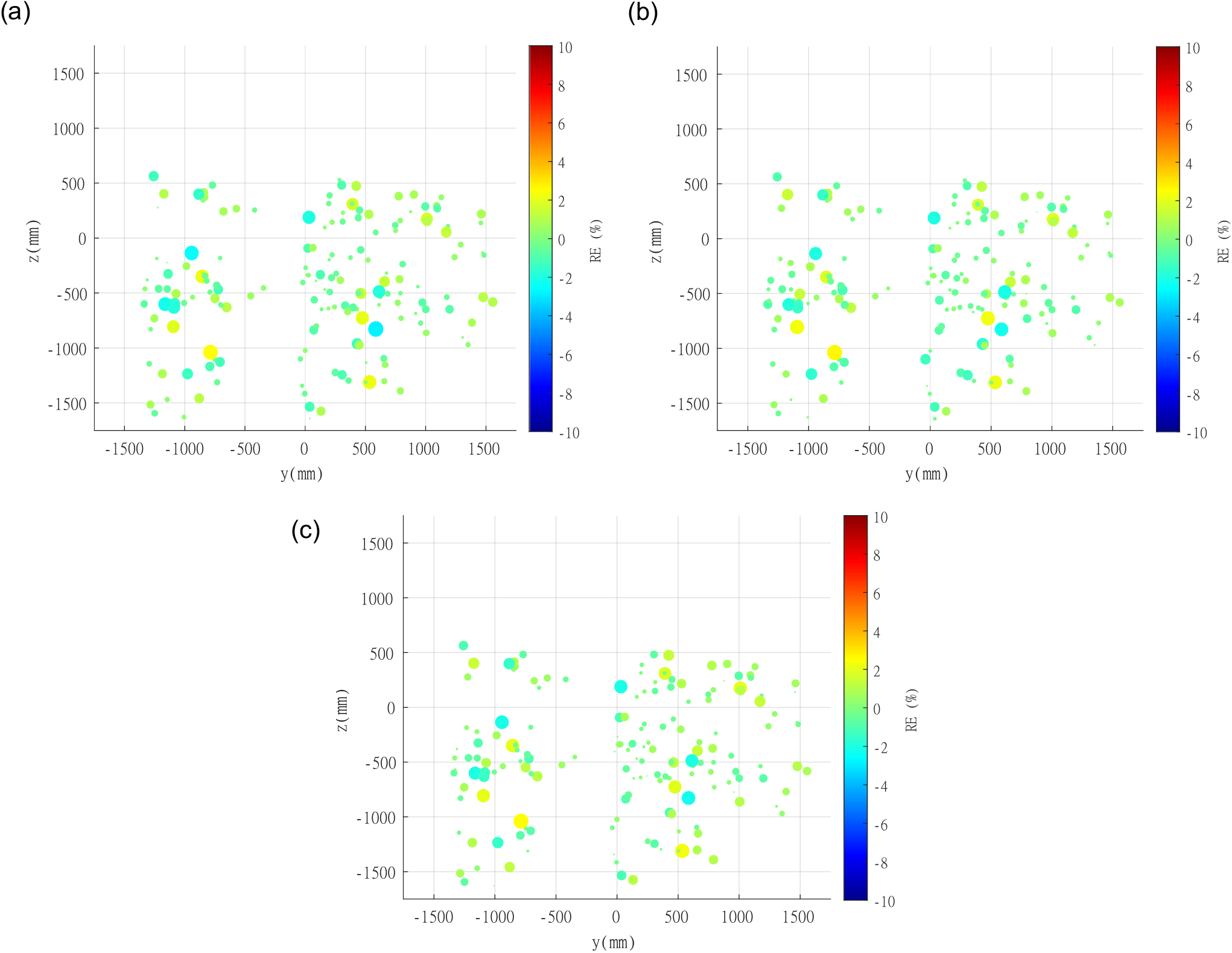}
\caption{Results: Percentage Splitting RE values \textit{y-z} plane comparison between models. Here, all models are trained on 90\% of the dataset Size of the circles are directly proportional to the magnitude of error. (a) Pure Data-Driven results (b) Sequential hybrid results (c) OPTMA-Net results. 
}
\label{fig:PercError}
\end{figure*}

All models are run 5 times for each testing case, and their mean values are reported. Table \ref{tab:res_main} shows the result comparison for the Percentage testing between pure data driven models, sequential hybrid physics infused models, and OPTMA-Net. The training dataset is varied from 10\% up to 90\% of the total dataset. As mentioned before, this is a good way of testing a model's generalizability. OPTMA-Net performs better than a pure data driven model and a sequential hybrid physics-infused model for 90\% and 30\% training cases. For the remaining cases, the error is of the same order as the pure data driven model, with the sequential hybrid model having higher error values for the test case with 10\% training data. Figure \ref{fig:PercError} shows the 2D scatter plot containing the RE values in the y-z plane for the 90\% training case. The circle sizes are proportional to the magnitude of RE. This is a good way to visualize the error spread over the 2D plane.

\begin{figure*}
\centering
\includegraphics[keepaspectratio=true,scale=0.79]{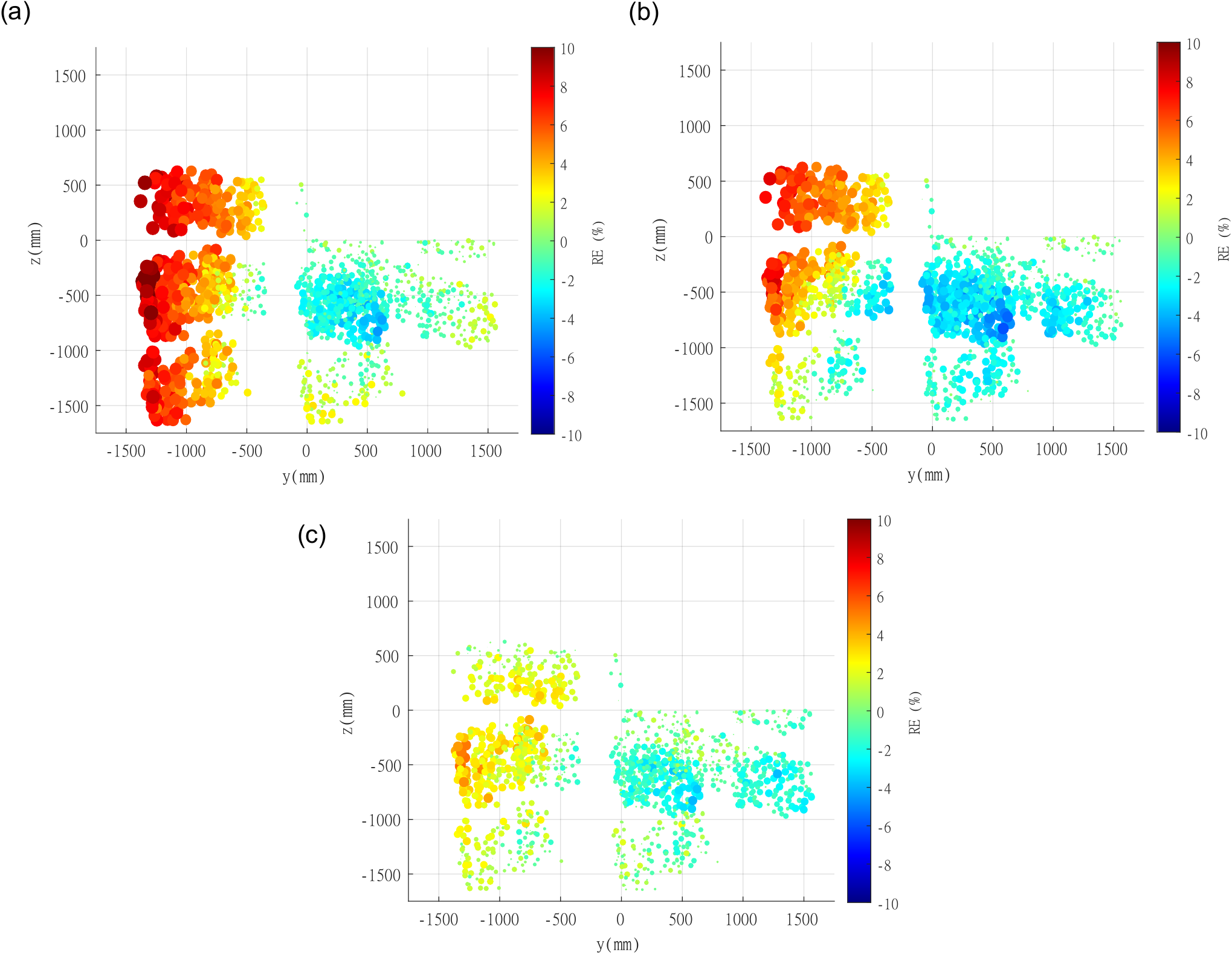}
\caption{Results: Quadrant Splitting RE values in the \textit{y-z} plane comparison between models. Here, all models are trained on 90\% of the dataset Size of the circles are directly proportional to the magnitude of error. (a) Pure Data-Driven results (b) Sequential hybridre sults (c) OPTMA-Net results.}
\label{fig:QuadError}
\end{figure*}

Unlike percentage testing, the quadrant testing scenario is much more difficult to predict for machine learning models as they are only trained based on the information from one quadrant. A prediction model must be able to handle the extrapolating requirements to have lower error values.  When tested on data points from unseen quadrants, OPTMA-Net has the lowest error values, which are reported in Table \ref{tab:res_main} over 5 runs. Pure data driven reports the highest error values for this testing. While performing better than a pure data-driven model, the sequential hybrid physics-infused model has higher error values than OPTMA-Net. Fig \ref{fig:QuadError} shows the error values in the y-z plane.

\begin{figure*}
\centering
\includegraphics[keepaspectratio=true,scale=0.79]{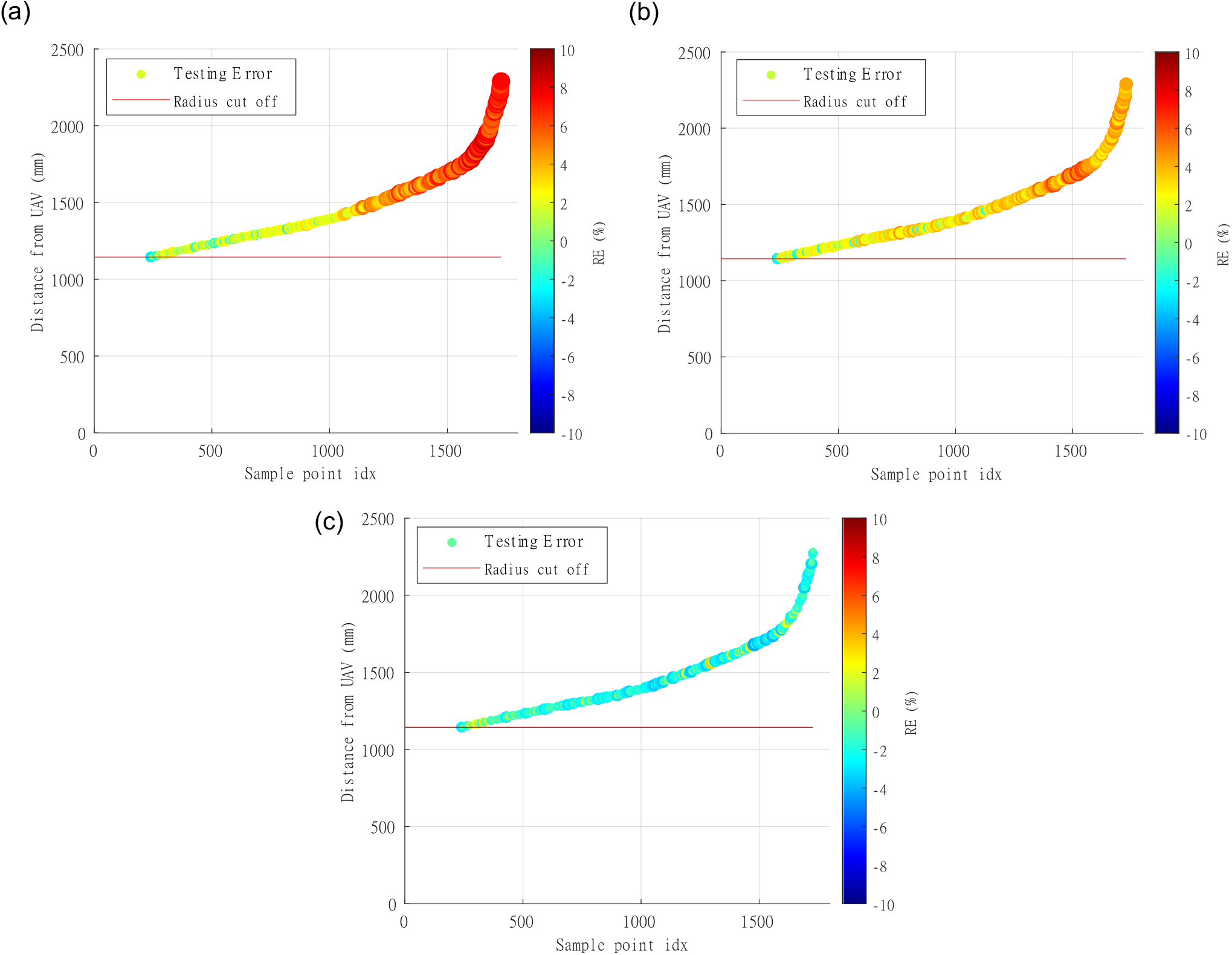}
\caption{Results: Radial Splitting RE values in the \textit{y-z} plane comparison between models. Here, all models are trained on 90\% of the dataset Size of the circles are directly proportional to the magnitude of error. (a) Pure Data-Driven results (b) Sequential hybridre sults (c) OPTMA-Net results.}
\label{fig:RadError}
\end{figure*}

The radial testing further measures the extrapolating capabilities of each model as the train-test data split is based solely on the total distance between a sample point and the UAV. Figure \ref{fig:Radial} shows how the splitting of training and testing data for this type of testing is done. Figure \ref{fig:RadError} shows the RE values of the testing dataset. Here, the error values are plotted against the radial distance of each point from the UAV. Just as we saw in quadrant testing, OPTMA-Net performs better than the other 2 models reporting an MSE  of 0.013 over 5 runs (Table \ref{tab:res_main}). Again the pure data driven model is seen to generate the highest error values among all 3 models.

\begin{table*}[t]
\begin{centering}
\caption{\textbf{Results: Normalized MSE on Test Samples}}
\makebox[\textwidth][c]{%
\begin{tabular}{ c c c c c c }
\toprule
Test Category & Training Data & Training Size & Pure DataDriven & Seq. Hybrid Model & OPTMA-Net\\
\hline 
\rowcolor{lightgray}
\multicolumn{6}{c}{Generalization}\\
\hline
Percentage Testing & 90\% & 1555 & 0.0083 & 0.0095 & \textbf{0.0083}\\
Percentage Testing & 70\% & 1209 & \textbf{0.0083} & 0.0089 & 0.0086\\
Percentage Testing & 50\% & 864 & \textbf{0.0083} & 0.0085 & 0.0083\\
Percentage Testing & 30\% & 518 & 0.0087 & 0.0091 & \textbf{0.0090}\\
Percentage Testing & 10\% & 172 & \textbf{0.0095} & 0.0239 & 0.0097\\
\hline
\rowcolor{lightgray}
\multicolumn{6}{c}{Extrapolation}\\
\hline

Quadrant Testing & $1^{st}$ Quadrant & 347 & 0.129 & 0.089 & \textbf{0.034}\\
\hline
Radial Testing & Sphere rad=114cm & 238 & 0.055 & 0.041 & \textbf{0.013}\\
\bottomrule
\end{tabular}
}
\label{tab:res_main}
\end{centering}
\end{table*}
%




%
%


%
%


%
%

\section{Conclusion}\label{sec:Conc}
In conclusion, a Physics-Infused Machine Learning architecture (OPTMA-Net) is presented here to enhance the extrapolation capability when predicting the sound pressure level due to a hovering quadcopter UAV at unseen locations. To test the extrapolation capability of the model, the model is trained and tested on data by using two segmentation techniques - 1) Split radially, and 2) Split into different quadrants. The aim with these data segmentation techniques is to create models that are able to provide promising accuracy even when predicting the Sound Pressure Level values at locations that belong to a different spatial distribution than the training locations. This is necessary as the acoustic field around a hovering drone is asymmetric and requires a model that captures the asymmetry of the system which is developed with the understanding of the partial physics of the problem.
The results highlight the problems associated with pure data driven models and their black box nature while showcasing OPTMA-Net's extrapolating and generalizing capability. While a black box neural network is able to predict the percentage testing cases with high accuracy, we see high MSE in both quadrant and radial testing. In quadrant testing, high errors are seen in the testing quadrants as they fail to capture the underlying physics-based trends and suffer from the black box prediction nature. The sequential hybrid physics-infused model gives better results than the data driven model but is not able to completely exploit the partial physics model. With that model we observed regions with high relative error. Further, the sequential hybrid model gives a high MSE of 0.0239 when trained with only 10\% of the total dataset. 

OPTMA-Net is observed to be roughly 4-fold more accurate than a pure data driven model, and well over 2-fold more accurate than the sequential hybrid physics-infused model, in the extrapolation cases -- i.e., for quadrant and radial data segmentation cases. As the partial physics model is included in the neural network structure itself, it provides a unified network with retrievable layers. This makes OPTMA-Net easy to run since the outputs do not have to be separately passed through the partial physics model, unlike in the original OPTMA architecture. It also enables physics-based analysis of the results by looking at intermediate layer values and their real-world relevance to the partial physics model. Since the model is written in PyTorch compatible terms, OPTMA-Net benefits from utilizing PyTorch's efficient and computationally fast training with back-propagation using AD. Such an architecture implementation lays the foundation to be readily adopted in modeling (at unprecedented accuracy and efficiency) a wide variety of system behavior, as long as low-fidelity partial physics models are also available to describe that behavior. To  promote such wider applicability, there remains notable scope of improvement for OPTMA. These include the ability to systematically identify the transfer features to be estimated by the transfer network, explore the possibility of substituting the partial physics model itself with a neural network trained on a dense set of partial physics samples, and allow the incorporation of multi-fidelity partial physics models.

\section*{Acknowledgments}

Assistance by Chen Zeng in operating the UAV during testing is gratefully acknowledged. Partial support of this work by the "SMART Start Phase II" program, from the Sustainable Manufacturing \& Advanced Robotic Technologies Center of Excellence at the University at Buffalo is also gratefully acknowledged. Support from the DARPA Award HR00111890037 is gratefully acknowledged. Any opinions, findings, conclusions, or recommendations expressed in this paper are those of the authors and do not necessarily reflect the views of the DARPA.
\newpage
\clearpage
\bibliographystyle{unsrt}
\bibliography{asme2e}
\end{document}